\documentclass{article}
\usepackage{graphicx}
\usepackage[round]{natbib}
\usepackage{epsfig}
\title{Five year prediction of Sea Surface Temperature in the Tropical
Atlantic: a comparison of simple statistical methods}

\author{
Thomas Laepple (AWI)\\
Stephen Jewson (RMS)\footnote{\emph{Correspondence email}: \texttt{stephen.jewson@rms.com}}\\
Jonathan Meagher (NOAA)\\
Adam O'Shay (RMS)\\
Jeremy Penzer (LSE)}

\begin{document}
\maketitle

\begin{abstract}
We are developing schemes that predict future hurricane numbers by
first predicting future sea surface temperatures (SSTs), and then
apply the observed statistical relationship between SST and
hurricane numbers. As part of this overall goal, in this study we
compare the historical performance of three simple statistical methods for making five-year SST
forecasts. We also present
SST forecasts for 2006-2010 using these methods and compare them
to forecasts made from two structural time series models.
\end{abstract}

\section{Introduction}

The number of hurricanes occurring in the Atlantic Ocean basin has increased in recent years, and
this has led to considerable interest in trying to predict future levels of hurricane activity.
One sector of society that is particularly interested in the number of hurricanes that may occur in the future
is the insurance industry, which pays out large amounts of money when severe hurricanes
make landfall in the US. The timescales over which this industry is most interested
in forecasts of hurricane activity are, roughly speaking, a zero-to-two year timescale,
for underwriters to set appropriate insurance rates, and a zero-to-five year timescale,
to allow financial planners to ensure that their business has sufficient capital to withstand
potential losses.

Motivated by this, we are in the process of building a
set of models for the prediction
of future hurricane numbers over these timescales.
The models in our set are based on different methodologies
and assumptions, in an attempt to understand how different methodologies and
assumptions can impact the ultimate predictions. Within the set, one subset of methods is based
on the idea of first predicting sea surface temperatures (SSTs), and then predicting hurricane
numbers as a function of the predicted SSTs.
The rationale for this
approach is that there is a clear correlation between SST and hurricane numbers, such that
greater numbers of hurricanes occur in years with warmer SSTs.
How, then, should we predict SSTs in order to make hurricane number predictions on this basis?

\citet{j92} compared three simple statistical methods
for the \emph{one-year} forecasting of tropical Atlantic SST. Their
results show that the relative skill levels of the forecasts
produced by the different methods they consider is determined by a
trade-off between bias and variance. Bias can be reduced by using a two
parameter trend prediction model, but a one parameter model that ignores
the trend has lower variance and ultimately gives better predictions when
skill is measured using mean square error.
How are these results likely to change as we move from considering one-year
forecasts to considering five-year forecasts?
For five year forecasts both bias and variance are likely to increase,
but not necessarily in the same way,
and as a result which model performs best might be expected to change
compared to the results of~\citet{j92}. We therefore
extend their study to investigate which methods and parameter sets
perform best for five year predictions.

We also consider 2 new statistical models, known as `local level' and `local linear' models.
These models are examples of so-called \emph{structural time-series models} and are commonly
used in Econometrics.
We produce SST forecasts using these 2 additional methods,
and compare the forecasts with those from our original set of 3 methods.

\section{Data}

As in~\citet{j92} we use the SST
dataset HadISST~\citep{hadisst}, which contains monthly mean
SSTs from 1870 to 2005 on a 1$^o$x1$^o$ grid.
As in~\citet{j92}, we define a Main Development Region
SST index as the average of the SSTs in the region
(10$^o$-20$^o$N, 15$^o$-70$^o$W), although we differ from~\citet{j92}
in that we now use a July to September average rather than a June to
November average. This is because July to September SSTs show a
slightly higher correlation with annual hurricane numbers than the June to
November SSTs.

The HadISST data is not updated in real-time, and so to update this
dataset to the end of 2006 we use the NOAA Optimal Interpolation SST V2 data
which is available from 1981 to the present.
The July-September MDR index derived from the NOAA dataset
is highly correlated with that derived from HADISST
(with linear correlation coefficient of 0.98).

\section{Method}

Following~\citet{j92} we compare three simple
methods for predicting SST using backtesting on the MDR SST
timeseries.
\citet{j92} tested 1 year forecasts while we now test 1-5 year forecasts.

The basic 3 methods we use are:

\begin{enumerate}

\item Flat-line (FL): a trailing moving average

\item Linear trend (LT): a linear trend fitted to the data
and extrapolated to predict the next five years

\item Damped linear trend (DLT): An `optimal' combination of the flat-line and linear
trend (originally from~\citet{j58}).

\end{enumerate}

We compare predictions from these methods with predictions from two structural time series
prediction methods which are common in Econometrics (see for example~\citet{harveys93}).
These models are:
\begin{itemize}
    \item a \emph{local level} model, that assumes that the historic SST time series is a
    random walk plus noise.
    \item a \emph{local linear trend} model, that assumes that the historic SST time series
    is a random walk plus random walk trend plus noise
\end{itemize}

The local level model has two parameters (the amplitude of the random walk and the amplitude
of the noise) and captures the idea that the level of SST changes over time, but with some memory.
The local linear trend model has three parameters (the amplitude of the basic random walk, the amplitude
of the random walk for the trend, and the amplitude of noise) and additionally captures the idea
that SST is influenced by a slowly changing trend. We fit the two structural time-series model
to the historical data using maximum likelihood.

\section{Results}

\subsection{Backtesting skill }

\begin{figure}[!hb]
  \begin{center}
    \scalebox{0.5}{\includegraphics{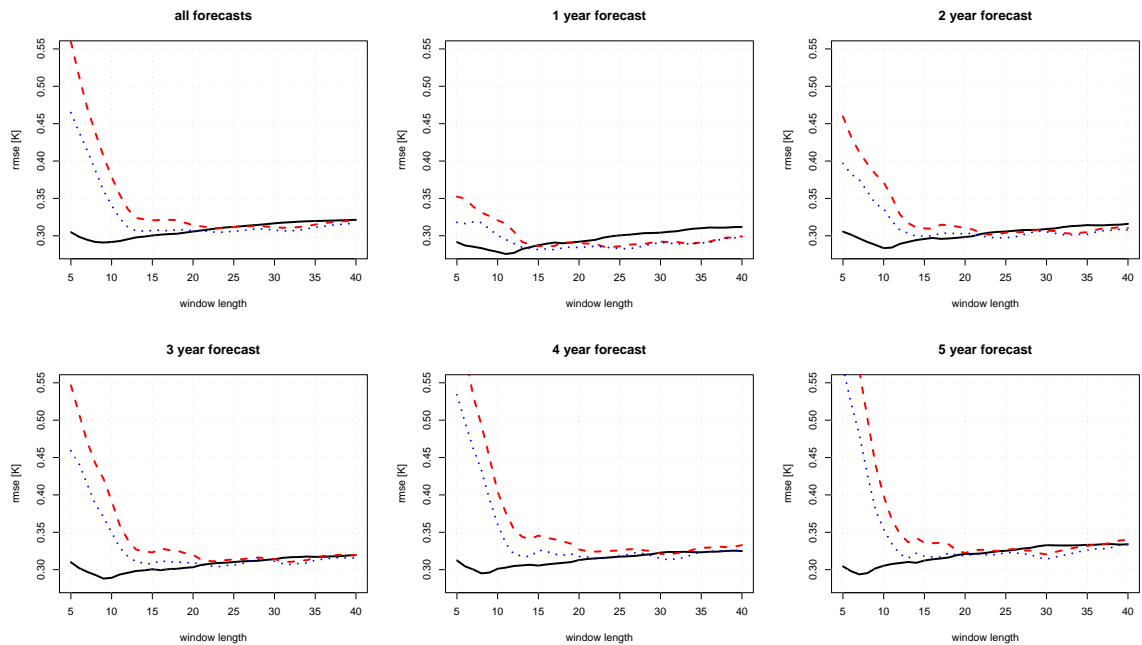}}
  \end{center}
\caption{forecast RMSE for the flat line model (black solid), linear
trend model (red dashed) and damped linear trend model (blue dotted) plotted
against the window length; the upper left panel shows the RMSE
over all forecast periods, the remaining panels show the RMSE for
specific forecast times.}
\label{f01}
\end{figure}

To compare the three basic prediction methods,
5 year periods from 1911-1915 to 2001-2005
were predicted (or `hindcasted') using from 5 to 40 years of prior data. Figure~\ref{f01} shows
the RMS error for all three models versus the number of years of prior data used.
The upper left panel shows the
score for 5-year forecasts, and the
other five panels show the scores for separate forecasts for 1 to 5 years ahead.

Considering first the RMSE score for the 5-year forecast, we see that the flat-line
model with a window length of 8-10 years performs best. Next best is
the damped linear trend model for a window length of around 17 years.
Worst of the three models is the linear trend model, which has an
optimal window length of 24 years.
The damped linear trend and linear trend models do very badly for short window lengths,
because of the huge uncertainty in the trend parameters when estimated
using so little data. Their performance
is then very stable for window lengths longer than 13 years.

We now consider the forecasts for the individual years.
First we note that the RMSE scores of these forecasts are scarcely lower than
the RMSE score for the 5 year forecast. This is presumably because the ability
of our simple methods to predict SST comes from the representation of long
time-scale processes. Our methods do not capture any interannual time-scale
phenomena.
Second, we note that the optimal window length for the flat-line forecast gradually
reduces from 11 years to 7 years as the lead time increases. This is the expected
behaviour of the flat-line model when used to model data with a weak trend.

\begin{figure}[!hb]
  \begin{center}
    \scalebox{0.5}{\includegraphics{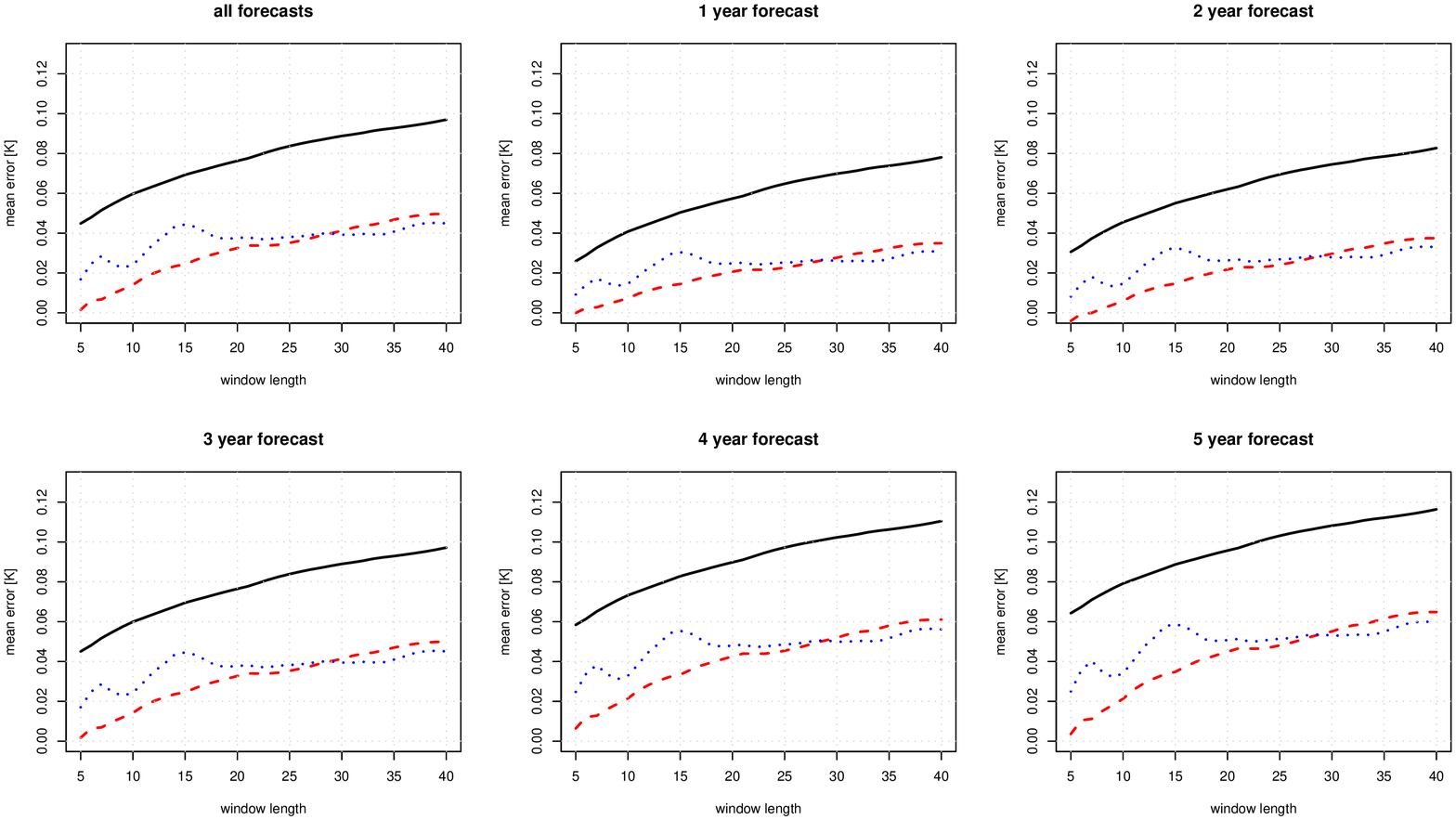}}
  \end{center}
\caption{forecast bias for the flat line model (black solid), linear
trend model (red dashed) and damped linear trend model (blue dotted) plotted
against the window length; the upper left panel shows the mean
bias over all forecast periods, the remaining panels show the bias
for specific forecast times.}
\label{f02}
\end{figure}

\begin{figure}[!hb]
  \begin{center}
    \scalebox{0.5}{\includegraphics{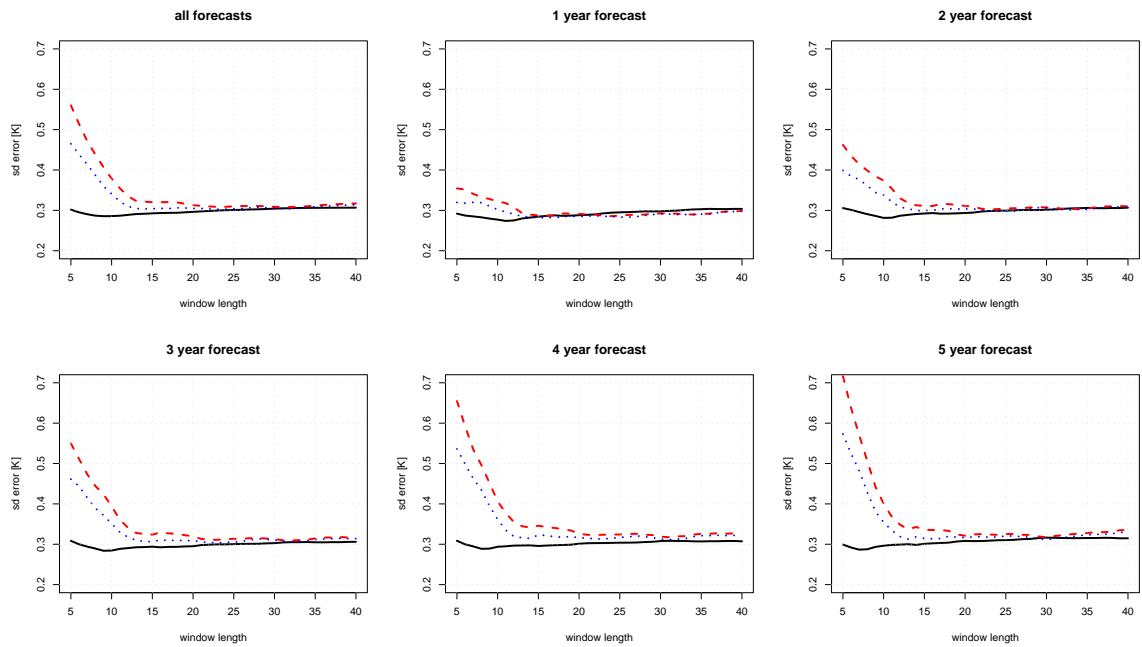}}
  \end{center}
\caption{standard deviation of the forecast error for the flat
line model (black solid), linear trend model (red dashed) and damped linear
trend model (blue dotted) plotted against the window length; the upper
left panel shows the SD error calculated over all hindcasts and
forecast periods, the remaining panels show the SD error for
specific forecast times.}
\label{f03}
\end{figure}

To better understand the error behaviour of these prediction methods we decompose
the RMSE into the bias and the standard
deviation of the error. Figure~\ref{f02} shows the bias for the three
models and figure~\ref{f03} their standard deviations. The flat line model
shows a high bias which increases with the averaging period
and the lead time. This is because using a flat-line cannot capture the trends in the
data.
%The bias in the linear model shows a similar behaviour
%but has a smaller increase than the flat-line model as a function of
%lead time because it can predict some part of the SST
%development during the five years of the forecast period.
%One would expect to see the damped linear trend
%model bias in-between the other two models and this is the case
%when looking at the bias for specific time periods (not shown).

Figure~\ref{f03} shows that it is the high variance in the predictions from the linear trend and
damped linear trend models, presumably due to high parameter uncertainty, which is
responsible for their poor performance when using small windows.
The standard deviation of the
flat line model error is close to independent of the lead time
although we can see that the minimum is shifted to smaller window lengths for longer
forecasts.

\subsection{Sensitivity of the results to the hindcast period}

\begin{figure}[!hb]
  \begin{center}
    \scalebox{0.5}{\includegraphics{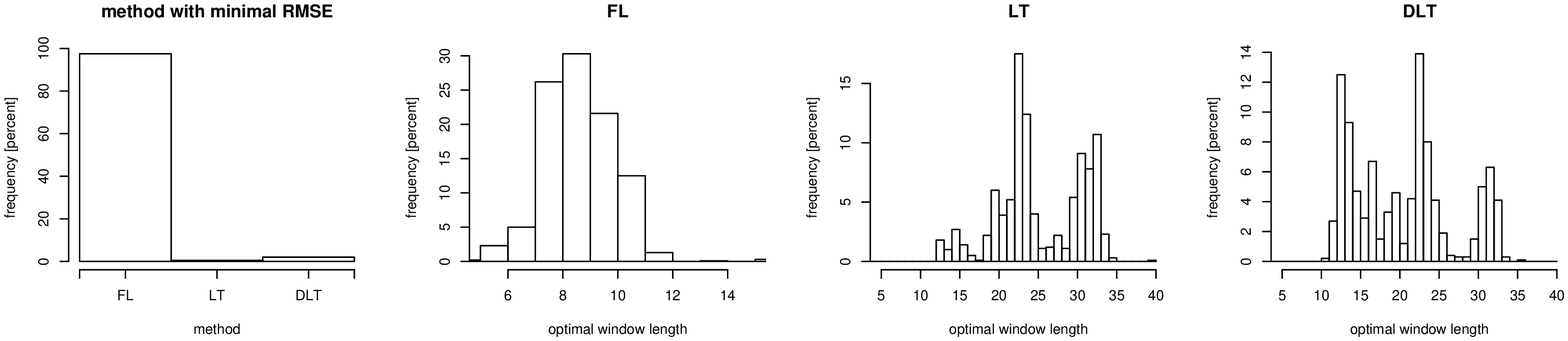}}
  \end{center}
\caption{sensitivity to the hindcast period for 5yr forecasts as
determined by bootstrap.}
From left to right; percentage of hindcast year samples in which a
specific method performed the best, distribution of optimal window
lengths for the flat line method, linear trend method and damped
linear trend method.
\label{f04}
\end{figure}

One obvious question concerns the stability of our results with respect
to the hindcast data we have used.
Understanding this should give us some indication of the
robustness of future forecasts. To check this stability we apply a bootstrap
technique by calculating the window-length dependent RMSE on
bootstrap samples of forecast years. Figure~\ref{f04} shows the results
for the five year forecast based on 1000 bootstrap samples. The
left panel shows the frequency in which one method outperforms the
other two methods, and the other panels show the distribution of optimal
window lengths for the three methods. For a five year forecast the
flat line method with a window length of 8 years is the best in almost all cases.
In contrast, the optimal window length of the linear methods
is strongly dependent on the hindcast years used.  However we note
that this is not necessarily a problem since the minima in the RMSE score
for these methods is very shallow and therefore an imperfect window length
does not greatly reduce the forecast quality.

\begin{figure}[!hb]
  \begin{center}
    \scalebox{0.5}{\includegraphics{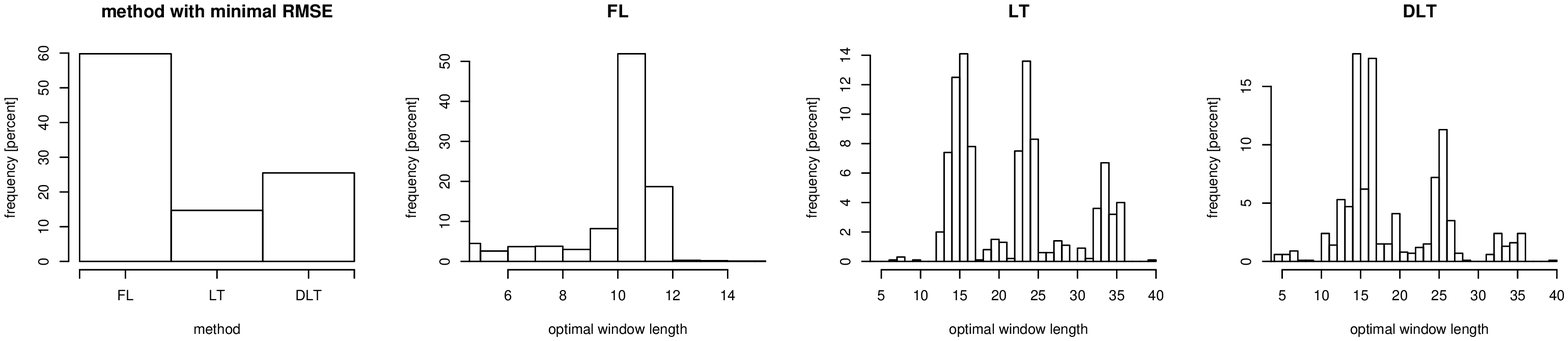}}
  \end{center}
\caption{sensitivity to the hindcast period for 1yr forecasts as
determined by bootstrap.}
From left to right; percentage of hindcast year samples in which a
specific method performed the best, distribution of optimal window
lengths for the flat line method, linear trend method and damped
linear trend method.
\label{f05}
\end{figure}

Figure~\ref{f05} shows the same experiment as the previous figure, but for
a one year ahead forecast. Here the linear trend models outperform
the flat line model in 40\% of the bootstrap samples and the
optimal window length of the flat line method is around 10 years,
confirming the results given in~\citet{j92}.

\subsection{Forecast for 2006-2010 and comparison to structural time series model
forecasts}

We now make forecasts for SST for the period 2006-2010 using the methods
described above.
Based on the backtesting results we use the flat line model with an 8 year window length,
the linear trend model with a 24 year window and the damped linear trend model with a 17 year window.

In addition we make forecasts with the local level and local linear structural time series models.
Point predictions from these models are the same as predictions from ARIMA(0,1,1) and ARIMA(0,2,2) models,
although predicted error distributions are different.

\begin{figure}[!hb]
  \begin{center}
    \scalebox{0.85}{\includegraphics{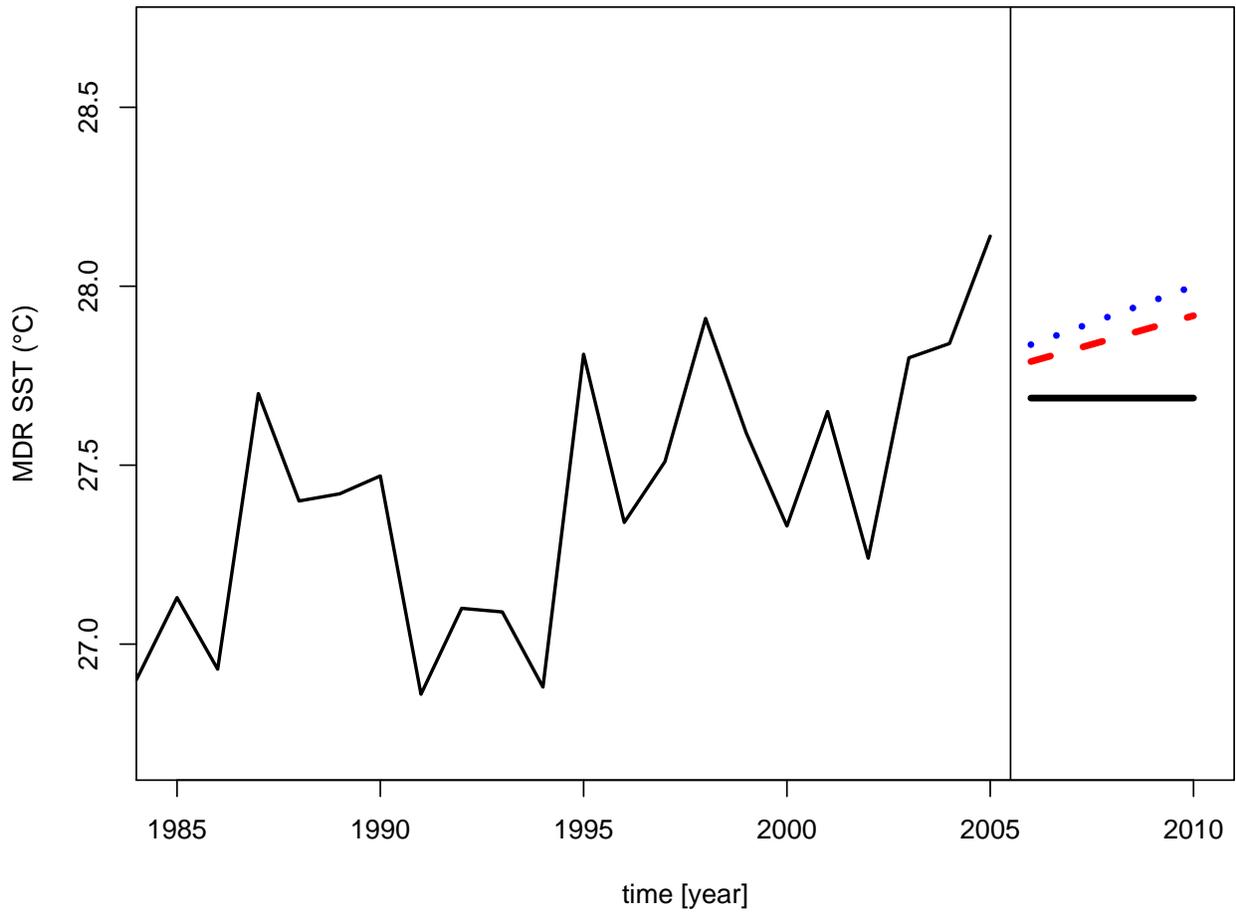}}
  \end{center}
\caption{Comparison of the 3 simple statistical forecasts for 2006-2010 and
their predicted RMSE. Flat-line (solid), linear trend (dashed) and damped linear trend (dotted).}
\label{f06}
\end{figure}

\begin{figure}[!hb]
  \begin{center}
    \scalebox{0.85}{\includegraphics{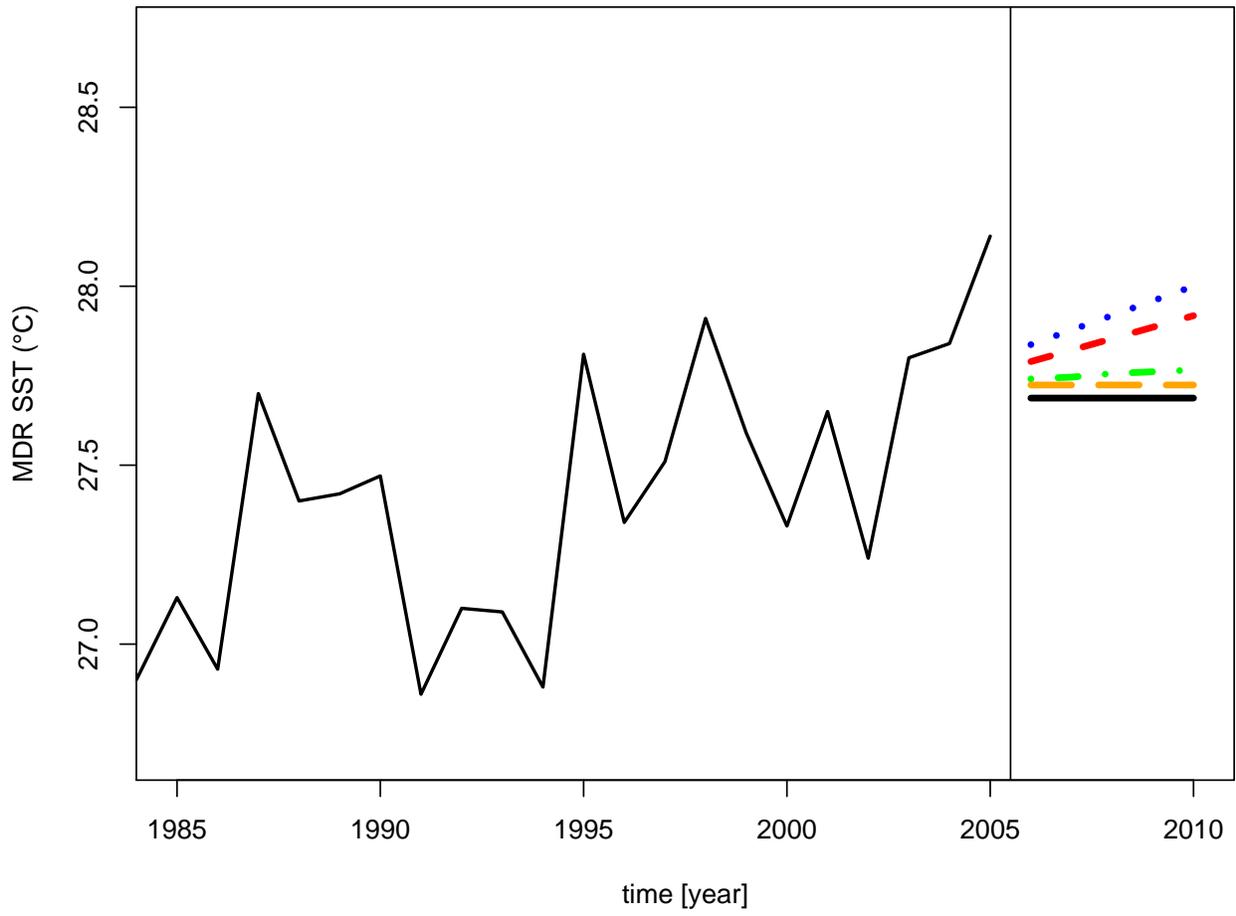}}
  \end{center}
\caption{
As in figure~\ref{f06}, but including predictions from the local level (long dashes)
and local linear (dot-dashed) models.}
\label{f07}
\end{figure}

\begin{figure}[!hb]
  \begin{center}
    \scalebox{0.85}{\includegraphics{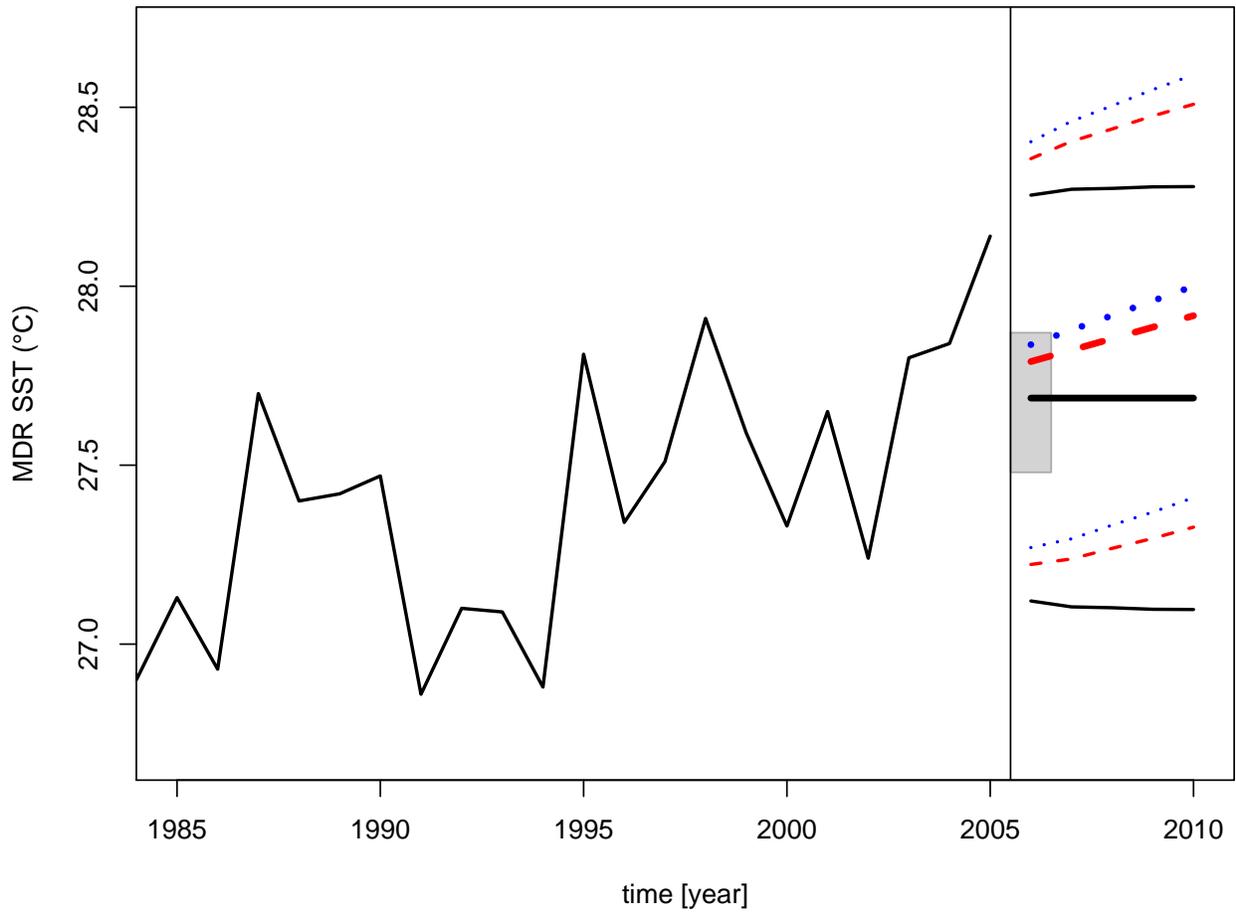}}
  \end{center}
\caption{
As in figure~\ref{f06}, but including (a) error bars showing plus/minus 1 standard deviation
and (b) a forecast for 2006, with 90\% confidence interval (grey box).}
\label{f08}
\end{figure}

Figure~\ref{f06} shows the forecasts from the 3 simple methods, not including the structural models.
As expected the linear trend
models predict higher SSTs than the flat-line models. Curiously, the damped linear trend model
actually predicts higher SSTs and a greater trend slope than the linear trend model. This
is because it uses a shorter window length than the linear trend model.
This unexpected behaviour slightly calls into question the way the damped linear trend model
is constructed, and suggests that there may be other ways that one could construct such an optimal
combination that might avoid this slightly awkward result.
It also highlights
the fact that the optimal window length for the linear trend models is not terribly well determined by the backtesting.
Figure~\ref{f07} also shows the predictions from the structural models. We see that these
predictions lie between the predictions from the flat-line and linear trend models.
Figure~\ref{f08} shows the predictions from the 3 simple models, but now including
(a) predicted RMSE scores for each model based on the backtesting results, and (b)
a prediction for 2006 based on data up to the end of 2005. To estimate the 2006 MDR SST
data we predict the July-September SST for 2006 using a linear
model with the NOAA Optimal Interpolation SST July-August data as
predictor ($1981-2005:R^{2}=0.913$). This point forecast and 90\%
confidence intervals are plotted in the figure as a grey box.

\section{Discussion}

We have tested a number of simple statistical prediction schemes
on historical SST data for the tropical Atlantic to evaluate their
forecast quality for a five year ahead SST forecast.
Our results are similar to those of~\citet{j92}, who tested the same
prediction methods for year-ahead forecasting.
The flat line method, a trailing
moving average, performed best using a window length of 8 years,
which is slightly lower than the optimal window length for year-ahead
forecasts. Next best was the damped linear trend method with window
lengths around 17 years. The linear trend method shows no advantage
over flat-line and damped linear trend for any forecast periods or window length.
By applying the hindcast experiment on subsets of hindcast data we
have shown that for the five year forecast the flat line methods
nearly always outperform the linear trend methods whereas for
a one year ahead forecast the linear methods are sometimes more
accurate.

It is worth remarking that the five year ahead forecasts we have
described have only around 10\% higher uncertainty than the one year ahead
forecast. It is likely that the one year ahead forecast
can be improved significantly by including additional information
such as the ENSO state, but for the five year ahead forecast the simple methods
we have presented will be more difficult to beat.

We have presented 5 year forecasts from both these simple methods and
local level and local linear trend structural time series models.
The forecasts from these structural time-series methods lie in-between the flat line and
linear trend forecasts and this suggests that one might consider the
flat line and linear trend forecasts as lower and upper bounds.

One final but important point is that our backtesting study has compared
the performance of forecast methods on average over the historical data.
Are methods that have worked well over the period covered by the historical
data likely to work well in the future? Not necessarily, since we seem to be in
a period of rapid warming. Although there are similar periods of rapid warming
in the historical data, there are also periods of cooling, and our backtesting
results reflect some kind of average performance over the two. If we believe that
the current warming will continue, then the methods that incorporate information
about the trend may do better than they have done over the historical
period, and the methods that ignore the trend may do worse than they have
done over the historical period.

\bibliography{arxiv}

\newpage

\newpage

\newpage

\newpage

\newpage

\newpage

\end{document}